# Large time-dependent coercivity and resistivity modification under sustained voltage application in a Pt/Co/AlOx/Pt junction

A. van den Brink,[1,a)] M.A.J. van der Heijden,[1)] H.J.M. Swagten,[1)] and B. Koopmans[1)]

[1] *Eindhoven University of Technology, PO Box 513, 5600 MB Eindhoven, The Netherlands.*

The coercivity and resistivity of a Pt/Co/AlOx/Pt junction are measured under sustained voltage application. High bias voltages of either polarity are determined to cause a strongly enhanced, reversible coercivity modification compared to low voltages. Time-resolved measurements show a logarithmic development of the coercive field in this regime, which continues over a period as long as thirty minutes. Furthermore, the resistance of the dielectric barrier is found to change strongly and reversibly on the same time scale, suggesting an electrochemical process is taking place within the dielectric. It is argued that the migration of oxygen vacancies at the magnet/oxide interface could explain both the resistance variation and the enhanced electric field effect at high voltages. A thermal fluctuation aftereffect model is applied to account for the observed logarithmic dependence.

(arXiv PrePrint Version, 29-10-2014)

[a] Author to whom correspondence should be addressed. Electronic mail: a.v.d.brink@tue.nl

## I. INTRODUCTION

The modification of magnetic anisotropy by application of electric fields across magnet-insulator interfaces has been experimentally observed in a variety of systems[1,2,3,4,5]. This method of magnetization control has enormous potential for spintronics applications, as the power consumption can be orders of magnitude lower than in current-based devices[6]. The magnitude, symmetry, and timescale of the observed effect differ significantly between experiments, suggesting that physics beyond the charging of an ideal capacitor play a role[7,8,9]. Notably, it has been proposed that the perpendicular magnetic anisotropy arising from Co-O hybridization at the magnet-insulator interface[10] can be directly affected by a voltage-induced migration of oxygen vacancies[11], resulting in strong voltage-induced anisotropy modifications[9]. Here, we measure the coercive field and leakage current as a function of time under sustained electric field, using Kerr microscopy on a perpendicularly magnetized Pt/Co/AlOx/Pt junction. The resulting data depends non-linearly on the bias voltage, and reveals a logarithmic time-dependence which we match to the hypothesis of electromigration of oxygen vacancies in the dielectric.

## II. SAMPLE FABRICATION AND EXPERIMENTAL SET-UP

Samples were fabricated on polished thermally oxidized silicon substrates using a lift-off electron-beam lithography procedure and DC sputtering at a base pressure of $1.0 \cdot 10^{-8}$ mbar, resulting in the structure depicted in Figure 1a. First, a 1.5 x 16 µm strip of Pt (4 nm) / Co (1.0 nm) / Al (2.1 nm) was deposited and exposed to an oxygen plasma at a pressure of $1.0 \cdot 10^{-1}$ mbar for 10 minutes. Subsequently, a 5 x 5 µm sheet of Al (2.0 nm) was deposited across the center of the strip and similarly oxidized, with the intent of insulating the side walls of the underlying strip. The total AlOx thickness after oxidation is estimated to be 6.0 nm[12]. Finally, a 1.5 x 16 µm strip of Pt (4 nm) was deposited orthogonally on top, enabling voltage application across the AlOx layers while still allowing for optical access to the underlying cobalt layer. Electrical access is provided through thick Ti/Au connecting electrodes (not shown).

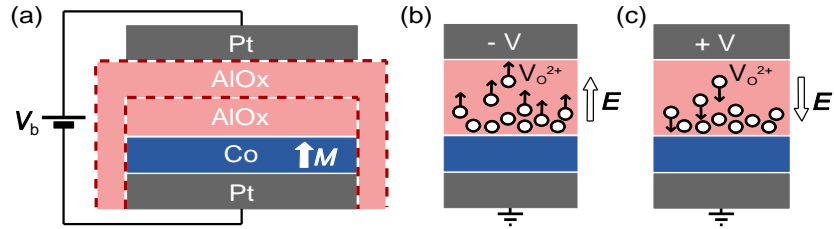

Figure 1: Schematics showing a) a cross-section of the fabricated junction, showing plasma oxidized interfaces as dashed lines, and the oxygen vacancy migration principle under b) negative or c) positive bias voltage.

The magnetization reversal process was studied using a Kerr microscope in polar mode, allowing for high-resolution digital imaging of the out-of-plane magnetization component at 16 frames per second. The magnetization was averaged over a 1 x 1 µm region of interest at the center of the junction. A differential method is used, where magnetic contrast is enhanced by subtracting a background image recorded at zero magnetic field before each experiment. A Keithley 2400 SourceMeter was used to apply voltages and measure the resulting current.

Immediately after deposition, magnetization reversal loops are recorded under application of a bias voltage. Starting at zero bias, the voltage is increased to higher positive and negative values in an alternating fashion, i.e. 0.0 V, 0.2 V, $-0.2$ V, 0.4 V, etc. At each voltage, the magnetization is first saturated by a $-50$ mT magnetic field perpendicular to the junction surface. Subsequently, the magnetic field is swept from $-30$ mT to $+30$ mT and back, resulting in two magnetization reversal events as reflected by changing intensity in the averaged Kerr image. This process (including saturation) takes three minutes, and is repeated several times to allow for averaging and to provide a time-resolved picture of the effect of a sustained voltage on the magnetic and electric properties of the junction.

## III. RESULTS

Typical hysteresis loops as obtained from the described experiment are depicted in Figure 2, showing a coercive field of approximately 10 mT at zero bias. The voltage-induced coercivity modification (panel a) shows two distinct regions: at low voltages (region I, between −0.6 V and +1.6 V) the effect is modest, at a slope of 0.5 mT/V. At higher voltages (region II) a much stronger coercivity modification is observed, at a slope of 4.5 mT/V. Furthermore, in region II the coercivity modification is found to increase with time under sustained electric field, as illustrated by panels b and d for an applied voltage of −2.0 V and +2.0 V, respectively.

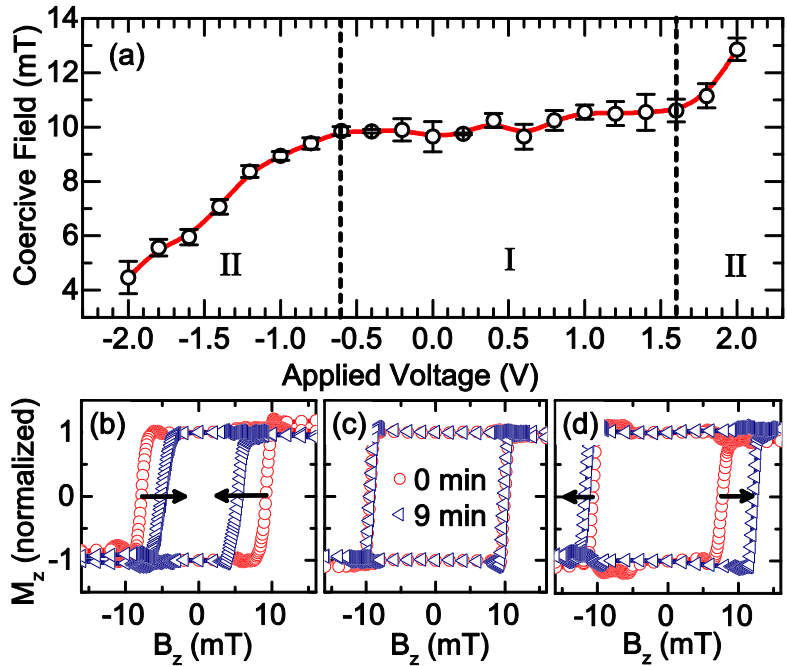

Figure 2: voltage-induced coercivity modification results. Panel a) shows the coercive field as a function of voltage, averaged over four measurements, identifying a low-sensitivity region I at low voltages and a high-sensitivity region II at high voltages. Note that the averaged values above ±1 V were computed using data recorded after twelve minutes of voltage application, to allow the voltage-induced effect to saturate sufficiently for a meaningful average to be computed. Typical magnetization reversal curves measured immediately after voltage application (red circles) and after 9 minutes (blue triangles) are shown for an applied voltage of b) -2.0 V, c) 0.2 V, and d) 2.0 V, with arrows emphasizing the evolution of the hysteresis loops over time. These plots have been corrected for linear drift and smoothed using nearest-neighbor averaging.

Having established that for higher bias voltages the properties of the junction evolve under sustained electric field, a more detailed study of the time-dependent coercivity modification and barrier conductivity was performed. Figure 3 shows the coercive field and leakage current as a function of time for a range of bias voltages. Before each measurement at negative voltage, the device was subjected to $+2.0$ V for 30 minutes to bring it into a relatively well-defined state, meaning that the voltage-induced effects are close to saturation. Similarly, a voltage of $-2.0$ V was applied for 30 minutes before each measurement at positive voltage. Hysteresis loops were then continuously recorded under sustained voltage application.

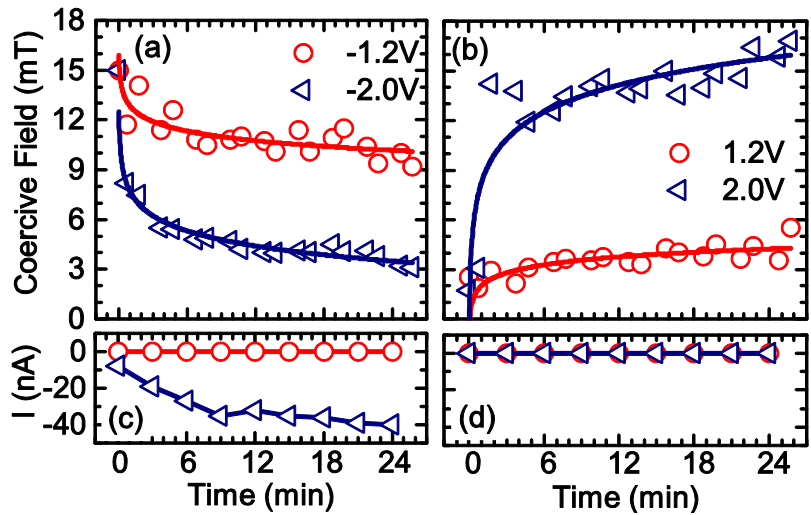

Figure 3: coercive field as a function of time for a) negative bias voltage, after applying +2.0 V for thirty minutes and b) positive bias voltage, after applying -2.0 V for thirty minutes. Logarithmic fits of the data are shown as solid lines. Note that the outliers at the start of the +2.0 V data are atypical; other measurements (e.g. at +1.8 V) follow the logarithmic trend more closely throughout. Simultaneously recorded leakage current values are shown for c) negative and d) positive bias.

The coercivity was found to decrease (increase) rapidly during the first few minutes of negative (positive) voltage application, continuously changing at a decreased rate during the remainder of the measurement, which lasted thirty minutes. The change is found to be reversible under application of an identical voltage of opposite polarity (compare $\pm 2.0$ V data) whereas a lower voltage of opposite polarity only partially reverses the effect (compare $\pm 1.2$ V data to $\mp 2.0$ V data). Furthermore, the obtained data was found to match a logarithmic trend, suggesting a thermally activated stochastic process as the physical cause of the effect. This hypothesis is explored in more detail below.

The leakage current flowing through the junction remained negligible (of the order of 0.01 nA) throughout the experiments, except under large negative bias voltage. For voltages larger than $-1.2$ V, the leakage current could be measured to increase over time, roughly following the logarithmic trend observed in the coercivity data. The final current magnitude, after thirty minutes of voltage application, was found to depend exponentially on the bias voltage. In all cases, reversing the bias voltage to $+2.0$ V caused a rapid decrease in the leakage current, reducing it to the original value of 0.01 nA in seconds.

## IV. DISCUSSION

The voltage-dependent coercivity data (Figure 2d) shows a small and apparently linear coercivity modification in region I at a slope of 0.5 mT/V, which can be explained by a modification of the interfacial anisotropy due to changing electron density, in line with existing experimental[1,2,3] and theoretical[13] work. Region II, by contrast, shows a strongly enhanced coercivity modification for bias voltages below $-0.6$ V and above $+1.6$ V. The symmetry around $+0.5$ V, matching the built-in voltage of the junction ($0.65 \pm 0.2$ eV[14]), implies that the effective electric field across the dielectric drives the enhanced coercivity modification. Furthermore, the timescale (minutes to hours) and the gradual and reversible nature of the observed effect suggest an electrochemical process rather than a purely electronic effect. A viable candidate for such a process is the electromigration of oxygen vacancy defects in the dielectric. This mechanism has been suggested to occur in similar devices[4,11,7,8] and was confirmed to affect the perpendicular magnetic anisotropy at Fe/MgO interfaces[11]. Moreover, electromigration of oxygen vacancies is known to occur in amorphous AlOx barriers[15], causing soft dielectric breakdown at high electric fields, similar to the behavior observed in Figure 3c.

The migration of oxygen vacancies under negative and positive bias voltage is sketched in Figure 1b and c, respectively, depicting an expected[16,17] filamentary distribution. A negative bias voltage then results in the gradual formation and growth of conductive filaments in the dielectric, whereas a positive bias voltage abruptly breaks existing filaments near the anode, in agreement with the observed leakage current data (Figure 3).

The observed logarithmic change of the coercivity under sustained voltage application can be explained using a thermal fluctuation aftereffect model[18]. Suppose a site $i$ at the Co/AlOx interface has a chance $P_i$ of being occupied by a charged defect, with an energy barrier $E_b$ separating it from a site that is $2E_{\text{hop}}$ lower in energy due to the applied electric field, then:

$$dP_i/dt \propto (1 - P_i)\exp[-(E_b + E_{\text{hop}})/k_B T] - P_i \exp[-(E_b - E_{\text{hop}})/k_B T], \quad (1)$$

with $k_B T$ the thermal energy. Assuming $P_i(t = 0) = 1$ and $P_i(t \to \infty) = 0$ this yields:

$$P_i(t) = 1 - \exp[-t/\tau_i], \quad (2)$$

with $\tau_i$ the site-specific relaxation rate, depending on the local energy landscape. If the relaxation rate is distributed uniformly across all sites, this can be shown[19] to result in:

$$\Delta n/n_0 = a + b \ln(t), \quad (3)$$

with $\Delta n/n_0$ the relative change in the charged defect density at the interface and $a$ and $b$ constants describing the instantaneous and time-dependent density variation, respectively.

The logarithmic time-dependence of the coercivity and resistance under sustained voltage application (Figure 3) can thus be accounted for by thermally activated electromigration of charged oxygen vacancies. Furthermore, this process also explains the observed symmetry around the built-in voltage, the asymmetric resistance variation, and the strong, non-linear, reversible nature of the coercivity modification at high bias voltages. Furthermore, the partial recovery of the coercivity under reduced reverse bias voltage (e.g. -1.2 V after +2.0 V, see Figure 3) agrees with the notion of a distribution of site-specific relaxation rates, allowing some defects to migrate relatively easily at low voltages.

The observed voltage-induced effects show both similarities and striking differences when compared to existing results. Using a Pt/Co/GdOx/Au system, Bauer *et al.*[4] demonstrated a large coercivity and resistivity modification close to the hard dielectric breakdown voltage. They identified migration of oxygen vacancies and charge trapping in the GdOx layer as possibly relevant physical mechanisms. In contrast to our observations, however, the effects Bauer *et al.* observed were irreversible and occurred for only one voltage polarity. Moreover, the authors suggest that the accumulation of oxygen vacancies decreases the interfacial anisotropy in their system, whereas our measurements indicate the opposite. These differences are likely to result from the different dielectric materials and oxidation profiles, both of which affect the soft dielectric breakdown process. A systematic study of electric field effects as a function of dielectric material and oxidation grade is called for, but is beyond the scope of this publication.

## V. SUMMARY

The coercivity of a Pt/Co/AlOx/Pt junction was found to decrease (increase) reversibly under negative (positive) bias voltage. For high bias voltages, this modification became strongly enhanced and time-dependent, following a logarithmic trend. High negative voltages were seen to cause an increased leakage current following a similar trend, while positive voltages instantaneously quenched it. These observations could be explained by the electromigration of charged oxygen vacancies near the Co/AlOx interface, driven by the effective electric field within the dielectric. A thermal fluctuation aftereffect model was used to account for the observed logarithmic time-dependence. Our findings shed more light on recent observations of enhanced electric field effects on large timescales, which may lead to ultra-low power spintronic devices for niche applications.

## ACKNOWLEDGEMENTS

This work is supported by NanoNextNL, a micro and nanotechnology programme of the Dutch government and 130 partners.